# A Survey: Implementations of Non-fungible Token System in Different Fields

A. N. M. Sajedul Alam[1], Junaid Bin Kibria[1], Al Hasib Mahamud[1], Arnob Kumar Dey[1], Hasan Muhammed Zahidul Amin[1], Md Sabbir Hossain[1], Annajiat Alim Rasel[1]

[1] Department of Computer Science and Engineering, School of Data and Sciences (SDS), BRAC University, 66 Mohakhali, Dhaka 1212, Bangladesh
{a.n.m.sajedul.alam, junaid.bin.kibria, al.hasib.mahamud, arnob.kumar.dey, muhammed.zahidul.amin, md.sabbir.hossain1}@g.bracu.ac.bd,
annajiat@bracu.ac.bd

**Abstract.** In the realm of digital art and collectibles, NFTs are sweeping the board. Because of the massive sales to a new crypto-audience, the livelihoods of digital artists are being transformed. It is no surprise that celebs are jumping on the bandwagon. It is a fact that NFTs can be used in multiple ways, including digital artwork such as animation, character design, digital painting, collection of selfies or vlogs, and many more digital entities. As a result, they may be used to signify the possession of any specific object, whether it be digital or physical. NFTs are digital tokens that may be used to indicate ownership of one-of-a-kind goods. For example, I can buy a shoe or T-shirt from any store, and then if the store provides me the same 3D model of that T-Shirt or shoe of the exact same design and color, it would be more connected with my feelings. They enable us to tokenize items such as artwork, valuables, and even real estate. NFTs can only be owned by one person at a time, and they are protected by the Ethereum blockchain — no one can alter the ownership record or create a new NFT. The word non-fungible can be used to describe items like your furniture, a song file, or your computer. It is impossible to substitute these goods with anything else because they each have their own distinct characteristics. The goal was to find all the existing implementations of Non-fungible Tokens in different fields of recent technology, so that an overall overview of future implementations of NFT can be found and how it can be used to enrich user experiences.

## 1 Introduction

To introduce an NFT (non-fungible token) it can be said that it is a unique piece of data that employs technology to log and authenticate digital information, such as digital artwork, 3D models, videos, songs, and images, on cryptocurrency blockchains, most notably Ethereum, which is itself a type of cryptocurrency. A "fungible" item is one that may be exchanged for another of the same value, such as a

$5 bill for another $5 bill. Cryptocurrencies are fungible because they employ a blockchain, which is a digital public ledger of transactions. An NFT is essentially a tradable code that is connected to metadata, such as an image, video or json or xml or text file, it can be any kind of metadata. The sale is recorded on a digital ledger (a blockchain) by a secure network of computers, giving the buyer proof of both legitimacy and ownership and which makes it more reliable to users for spending or investing their hard earned money on NFT trading.

In educational institutions students' evaluations and rewards are crucial parts of the educational process. One of the most efficient positive reinforcement tactics is gamification. In these systems, the learning process is turned into a game with a positive reward structure. Using Non-Fungible Tokens (NFTs) as an incentive mechanism related to student assessments, Elmessiry et al. proposed a novel way to positively incentivize both students and teachers [1].

Distance learning will become a primary answer for higher education organizations facing an unsustainable economic future as the global population seeks various sorts of education. Decreased physical infrastructure along with face-to-face interaction requirements will enable students from all over the world to overcome geographical restrictions that may prevent them from enrolling in on-campus courses [7].

They provided some strategies to increase the effectiveness of the learning and those are hybrid models for both online and offline, updated materials, proper engagement, adaptive teachers, engaging online students, initiating short answer exams and optimum course design. They have also touched on gamification learning where students will remain engaged in classes just like they are engaged in games. Including gamification in a revenue model that includes NFTs would transform the way people learn in both academic and skill-based settings. Institutions of higher learning that combine the cost-cutting qualities of hybrid training and education with the incentive structures provided by NFTs both to learners and educators will find a radically disruptive approach to knowledge transfer and learning.

The transaction made news, and NFTs have subsequently become extremely popular. In 2021, investors spent $27 billion into the market, and Meta, Facebook's renamed parent company, now aims to allow users to create and sell NFTs, according to reports. The only difficulty is that the NFT market will eventually fail, for a variety of reasons.

The NFT market is going to face a similar fate — but not due to environmental concerns, as some may believe. To be true, NFTs utilize a lot of energy since cryptocurrencies like Ethereum and Bitcoin are "mined" using massive networks of computers that have a large carbon footprint that grows with each transaction. However, climate change is a red herring when it comes to determining what will bring the NFT market down. The actual issue is that the present NFT craze is based on a shaky foundation.

To create a customized, distributed, and impartial basic model for trading games, Christos et al. used the InterPlanetary File System (IPFS) and Non-Fungible Tokens (NFTs) supported with Distributed Ledger Technologies (DLTs) [8]. Their approach builds a completely distributed system that enables innovative business models by allowing obtainable game materials to be traded and priced based on scarcity, even while providing a share to the digital artist, often without the requirement for a trusted intermediary. The technology ensures that properties stay on the internet, so players don't have to worry about missing ownership of their artworks or their worth if the game's developer gets bored or declares bankruptcy.

In current history, the Non Fungible Token (NFT) sector has already seen multi-million dollar trading. NFTs are one-of-a-kind and scarce digital treasures that are protected by a blockchain. An NFT ensures the originality of source, authority, rarity (scarcity), and long-term preservation of a certain object. Market microstructure,Alternatively, examines the mechanism through which traders' hidden desires are transformed into performed transactions. [10]

As NFTs have achieved remarkable attention from investors, the question arises about the value and scarcity of blockchain technology. Traditional currency systems are fungible objects. A non-fungible asset are those products which do not pay the same or identical price for interchange of products. The title-map of Etheria is considered as the first NFT which was dated to 2015, the majority of the titles remain unsold. The pixelated pictures of Cryptopunks was the second NFT in 2017 and the third one was in 2017, pixelated concept of MoonCatRescue. [6]

As we know, Non-Fungible Token (NFT) is sort of a cryptocurrency which differs from other classical cryptocurrencies such as Bitcoin. In industrial and scientific communities NFT has gathered remarkable attention as people have shown interest in different NFTs. In this article, authors provided core components, roadmap, and opportunities of NFTs by describing current NFTs designing solutions, security solutions, current opportunities, and open challenges of the NFTs ecosystem. [9]

The digital art market has been considerably growing thanks to the combination of the Blockchain technology and non-fungible tokens. NFTs attracted bil-lions of dollars in investment. Christie's, the British auction house founded in 1766 by James Christie, sold an NFT for 69 million dollars. The paper proposes a predictive tool, NFT price Oracle, as a solution for users that wish to understand the value of an artwork (and thus how much to pay for the NFT associated with the artwork) in the Blockchain. In Particular, it needs to be decentralized, to better guarantee that faked digital artworks are not paid as if they were original.

Pierro et al. [11] proposes a predictive tool, NFT price Oracle, as a solution for users that wish to understand the value of an artwork (and thus how much to pay for the NFT associated with the artwork) in the Blockchain. In Particular, it needs to be decentralized, to better guarantee that faked digital artworks are not paid as if they were original.

Genobank.io is a project that intends to preserve consumer privacy by bringing together players from the fields of privacy legislation, smart contracts, and genomics. The California Consumer Privacy Act (CCPA) and its ramifications for smart contracts, especially NFTs in the blockchain, are discussed. The vast majority of the world's genetic data has yet to be created. The usage of blockchain with NFTs (a sort of smart contract) might be subject to privacy restrictions. Several iterations of privacy-by-design services are being developed by businesses to solve this difficulty. [12]

Cryptocurrency is a digital asset that does not depend on banks or financial institutions to corroborate transactions. It's a peer-to-peer system secured by cryp-tography, which makes it nearly impossible to spend double. There are four kinds of blockchain networks —public, non-public, association and hybrid blockchains. A hybrid blockchain combines centralized and decentralized features. Quantum non-fungible tokens (NFT) are a type of cryptographic tokens that are extremely powerful with fundamental properties unique, traceable,rare, programmable, and indivisible.

The sale of NFTs was predicted to be worth $12 million in December 2020, but in February 2021, it soared to $340 million in just two months. Our experiment was carried out on a cloud-based quantum computing platform, i.e., IBM Quantum Experience [14], available on the Web.

A stake characterizes the esteem or money one tends to wager on a particular outcome, and the strategy is named staking. A validator stake is outlined by the product of the number of coins with the number of times a single user has controlled them.

Pandey et al. [13] discussed that weighted double hypergraph states can be created by adding phase to each qubit by local operation. Each qubit of class A and B are entangled in a bell state with each other at their respective vertices. The hypergraph's number of vertices breaks even with n, the double number of qubits within the quantum frame-work.

In this paper, we are surveying in total 8 papers to explore which implementations are already done of NFT and how we can use them in our future technology. The NFT market is growing really fast since the beginning of 2021 and in the future NFT will be a more valuable asset compared to the present. More and more people will eventually invest in it eventually. So, our goal is to show all the aspects of NFT in current technology by which people can understand the NFT and about the usage of NFT better. NFT not only created a huge business platform but also encouraged digital artists to work further. These are only two major aspects of NFT but not the end, there are thousands of possibilities within NFT which people barely even know. Our paper can be helpful for finding those possibilities and visions of NFT we believe.

## 2  Survey Details

Chohan [6] deals with the two questions about NFTs. The first question is about whether NFTs are really rare and the second question is about the ownership, does the owner really own it. Furthermore NFT cannot provide the actual location of the owner. The object which is represented by a token can be copied or the representation cannot be viewed explicitly. Hacking and theft of tokens can happen. NFT deploys surplus money to find its value for people. The value of NFT is not dependent on market or stocks of market, rather it is dependent on leisures of blockchain. NFT is a concept which offers interesting and creative use of tokens which may seem interesting to people.

Assessment and rewards for students are important aspects of the educational process. They are also important components in increasing student knowledge acquisition. Gamification is one of the most effective positive reinforcement strategies available. The learning process is made into a game with a positive reward system in these systems. Historically, due to the lack of supporting technology, building such a gamification system has been extremely challenging. Elmessiry et al. suggested a novel technique to positively incentivize both students and teachers by employing Non-Fungible Tokens (NFTs) as an incentive mechanism related to student assessments [1]. Their strategy provides for positive incentive reinforcement by giving

NFT holders accolades and privileged access to advantages. Decreased physical infrastructure along with face-to-face interaction requirements will enable students from all over the world to overcome geographical restrictions that may prevent them from enrolling in on-campus courses [7]. They provided some strategies to increase the effectiveness of the learning and those are hybrid models for both online and offline, updated materials, proper engagement, adaptive teachers, engaging online students, initiating short answer exams and optimum course design.

To create a customized, distributed, and impartial basic model for trading games, Christos et al. used the InterPlanetary File System (IPFS) and Non-Fungible Tokens (NFTs) supported with Distributed Ledger Technologies (DLTs) [8]. Their approach builds a completely distributed system that enables innovative business models by allowing obtainable game materials to be traded and priced based on scarcity, even while providing a share to the digital artist, often without the requirement for a trusted intermediary. The technology ensures that properties stay on the internet, so players don't have to worry about missing ownership of their artworks or their worth if the game's developer gets bored or declares bankruptcy.

Wang et al. [9] discussed the state of art NFT solutions and also discussed the protocols, standards, properties, opportunities of NFT. For NFTs protocols, NFTs require a ledger which stores records like a database to store NFT data. For NFT owner and NFT buyer, some characteristics are required like an NFT owner digitizing the raw data into proper format to check title, description and an NFT owner also storing the data into an external database which is outside of the blockchain. The miting process is completed when a transaction is properly done. For NFT, each of the tokens are unique than other tokens. A non fungible token standard differs from a fungible token.As NFTs are considered as decentralized, it requires verifiability, transparent execution, availability, usability, atomicity, and tradability. There are several fields which are getting benefits for NFTs like boosting gaming industry, digital collectibilities, virtual events, inspiring the metaverses etc.

Mukhopadhyay and Ghosh [10] discussed the microstructure of the NFT market, with an emphasis on price setting, industry structure, clarity, and applicability to other financial fields. Mainly the author divided their paper into four parts. They discussed price formation and discovery. The process through which prices constrain new knowledge is known as price formation.It is a technique of obtaining information to guarantee that market participants are sufficiently informed about the pricing of the NFTs exchanged in the marketplace. They also discussed Market Structure and Design. Market structure has an effect on the speed and quality of price discovery, as well as availability and trading costs. What counts in the end are the procedures which transform the investor's request into a completed transaction. NFT market stability and determination of health indicators are more essential from a design standpoint, rather than the amount of transactions in the NFT marketplace. There are two levels in the NFT marketplace. The first is at the project level, whereas the second is at the level of ownership. There are two types of information that influence market microstructure: transparency and disclosure. Market manipulation via Wash Trading must first be quantified by determining how Wash Trading occurs, how to spot it and what logical suspicions lead to deem a transaction an illegal one. In technical analysis, historical price movements are utilized to anticipate upcoming returns . Technical analysis may assist traders find hidden liquidity.

Pierro et al. [11] NFTs (non-fungible tokens) are digital assets that are both immutable and indestructible. NFTs also allow collectors to value digital art in the same manner that they value physical art, opening up new avenues for digital artists. The "Fake Bansky NFT " was sold for 336 thousand dollars on the artist's website. We suggest that there is a better, more user-centered way to assist NFT visitors in determining the ownership/authorship of artwork. The value of a specific NFT is determined by a range of circumstances, some of which are beyond its control.

The purpose of Uribe and Waters [12] is to guarantee internet privacy for clients on a nationwide basis. Individuals who have been harmed by a breach caused by a lack of security measures have a private right of action under the new law, which imposes fines ranging from $2,500 to $7,500 per violation [15]. Crypto-datawallets allow users to regain control of their genetic data. This would be the same as self-serving a consumer's right to be forgotten as a data subject/owner under GDPR and CCPA language. Adopting a crypto data wallet for genetic data has the primary purpose of providing users with a new alternative.

Genobank is the first (patented) personal DNA kit that ensures individuals have total ownership and control over their genetic information. Although the ERC-998 and future ERCs are better than prior ERCs, they still have issues such as inefficient transfer capability, array length, and the inability to access token IDs.

Pandey et al. [13] mentioned there is no physical data in NFT, such as paintings, videos, or photos. It only has information about the owner and the artwork (link of art or name of art). Each block can pass this level of verification due to the entanglement of the double weighted hypergraph state. Because entanglement maintains the security of the entire system, it cannot be tampered with. Each block is added to all copies of the chain instantly, making it user-friendly. The public is welcome to observe NFT operations such as mounting, purchasing, and selling.

## 3 Analysis

For analyzing our survey results, we have created the following Table 1 to provide a brief description of the key concepts of each article as well as categorizing them into sub-domains and sub-disseminations within each domain.

**Table 1.** The following is a list of the articles that were considered for this Systematic Literature Review.

| Articles | Major Domain | Sub-Domain | Key Concept |
| --- | --- | --- | --- |
| Market Microstructure of Non-Fungible Tokens [10] | Technical issues and solutions | Finance, Economy | This paper focuses on price formation, market structure, transparency, and applicability to other financial fields of the NFT market microstructure. Market manipulation in the NFT market has also been investigated in the context of wash-sale patterns. The essay finishes with recommendations for due-diligence activities that investors might use to reduce the danger of NFT trading. |
| Non-Fungible Token (NFT): Overview, Evaluation, Opportunities and, Challenges [9] | Overview and others | Value, Originality | Discussing the state-of-art of NFTs, this paper discusses properties, protocols, challenges, opportunities of NFT. The NFT ecosystem is explored based on systematic study in this paper. |
| Non-Fungible Tokens: Blockchains, Scarcity, and Value [6] | Technical issues and solutions | Value, Originality | Author has focused on ownership of NFTs and also whether NFTS are really rare. Questions have been raised based on value and scarcity with respect to blockchain technology with a target to enrich the path of blockchain development. |
| Original or Fake? How to Understand the Digital Artworks' Value in the Blockchain [11] | Technical issues and solutions | Value, Originality, Digital Artworks | Non-fungible tokens (NFTs) are certificates of ownership/authorship of digital work. Supported by the blockchain structure, these certificates are inviolable, unassailable, and indestructible. |

| | | | The paper proposes a predictive tool, NFT price Oracle, as a solution for users. |
|---|---|---|---|
| Privacy Laws, Genomic Data and Non-Fungible Tokens [12] | Policy and recommendation | Privacy and Laws, Genomic Data, Security, Copyright | Different legal frameworks apply to privacy laws, genomics, and smart contracts. This paper looks at how data engineers might overcome the difficulties of incorporating privacy regulations and other legal obligations into smart contracts. |
| Realizing non-fungible token (NFT) through IBM quantum experience [13] | Technical issues and solutions | Quantum Experience | A novel system for producing quantum non-fungible tokens (NFT) is presented in a study, in which a quantum state representing NFT is mounted on a blockchain instead of physically handing it over to the owner. Our protocol has the potential to replace traditional NFT and give clients a more secure and cost-effective option for item recognition. |
| NFT Student Teacher Incentive System (NFT-STIS) [1] | Overview and others | System, Application | The authors made a reward based system for both teacher and students so that the interaction during class remains high. They introduced gamification techniques and used non fungible tokens as an incentive. |
| Fully Decentralized Trading Games with Evolvable Characters using NFTs and IPFS [8] | Overview and others | System, Application | To create a flexible, decentralized, and fair system for trading games, we use the InterPlanetary File System (IPFS) and Non-Fungible Tokens (NFTs) supported by Distributed Ledger Technologies (DLTs). |

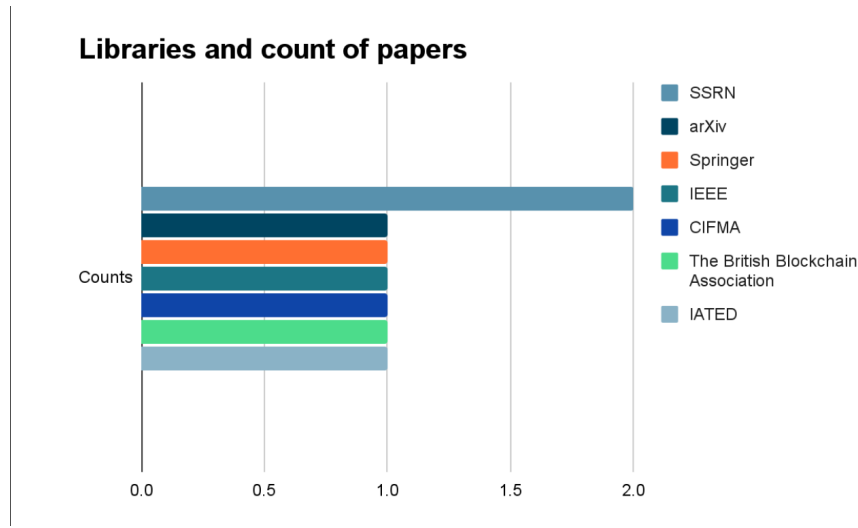

**Fig. 1.** Libraries and count of papers from them.

From **Fig. 1** the highest counting number indicates that SSRN library has most of the papers among the following papers were chosen. The other libraries such as IEEE, CIFMA, The British Blockchain Association, IATED, Springer, arXiv have the same amount of papers.

For this investigation, seven libraries were used to find interesting research subjects. This work picks certain conference papers from such publications' collections that are related to the study subjects that are engaged in a comprehensive literature review of non-fungible token technology acceptance. This study critically examines the papers linked to our study aim by employing the search capabilities of each publishing repository. To be even more exact, just the title, abstract and keywords are being used in the query. **Table 2** lists the libraries of all seven publishers.

**Table 2.** List of the libraries that were searched.

| No. | Library |
|---|---|
| 1 | SSRN |
| 2 | arXiv |
| 3 | Springer |
| 4 | IEEE |
| 5 | CIFMA |
| 6 | The British Blockchain Association |
| 7 | IAETD |

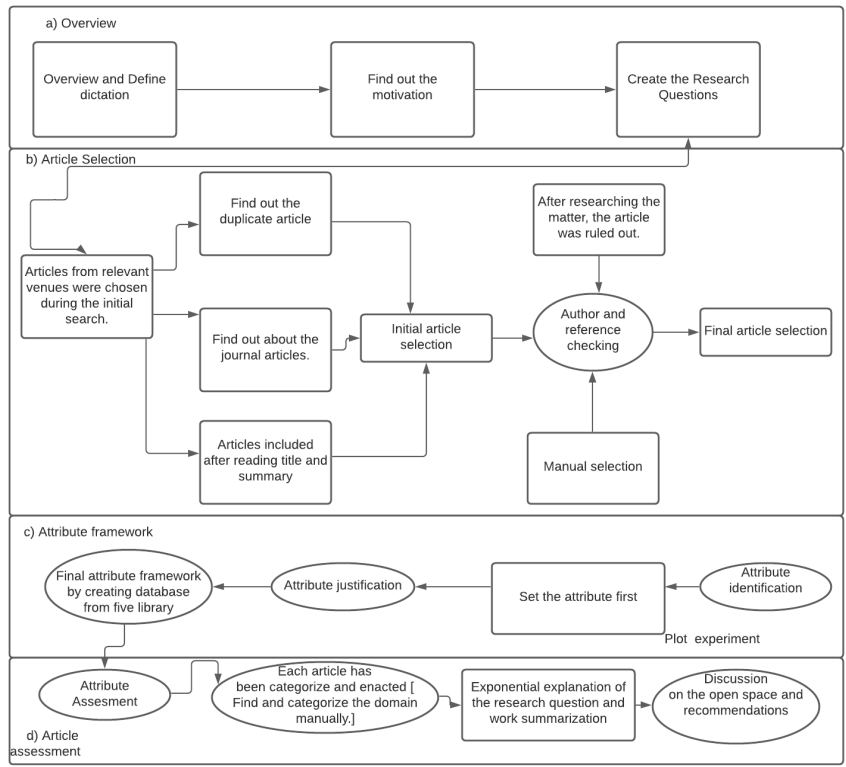

**Fig. 2.** Overview of Survey Details [2], [3], [4], [5]

The detailed description of survey details is proposed in **Fig. 2** which illustrates the beginning of the survey to end portion, that means how the articles were chosen, how the survey was done and how the final recommendations were found. The whole procedure of **Fig. 2** was chosen from Kitchenham's procedure which includes attribute selection, attribute framework, attribute assessment, research questions [2],[3], [4], [5].

## 3.1 Sub Domain List:

  I.   Value, Originality
  II.  System, Application

III. Digital Artworks
IV. Security, Copyright
V. Quantum Experience
VI. Genomic Data
VII. Laws and Privacy
VIII. Economy
IX. Finance

Based on these nine subdomains chosen by us, we have divided the eight papers into three major domains depending on three genres named, **'Technical issues and solutions'**, **'Policy and recommendations'** and **'Overview and others'**. We have shown which papers belong to which genre in **Table 3** and it can be seen that most of the papers belong to the **'Technical issues and solutions'** genre.

**Table 3.** Genre distribution of selected papers.

| Genres | Titles | Counts |
|---|---|---|
| **Technical issues and solutions** | a) Non-Fungible Tokens Blockchains, Scarcity, and Value<br>b) Original or Fake How to Understand the Digital Artworks' Value in the Blockchain<br>c) Realizing non-fungible token (NFT) through IBM quantum experience<br>d) Market Microstructure of Non Fungible Tokens | 4 |
| **Policy and recommendations** | a) Privacy Laws, Genomic Data and Non-Fungible Tokens | 1 |
| **Overview and others** | a) Non-Fungible Token NFT Overview Evaluation Opportunity<br>b) Fully Decentralized Trading Games with Evolvable Characters using NFTs and IPFS<br>c) NFT Student Teacher Incentive System (NFT-STIS) | 3 |

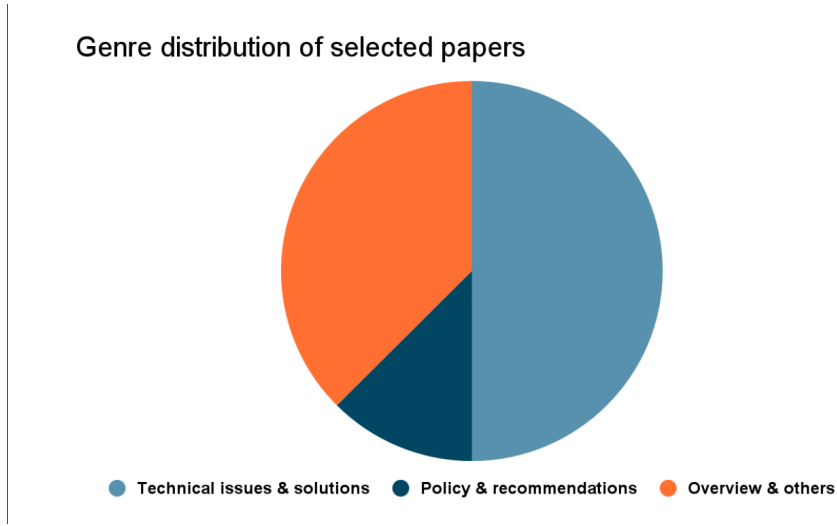

**Fig. 3.** Genre distribution of selected papers.

In **Fig. 3** we have demonstrated the genres by which we have divided our papers. In the categorization we found that half of our papers were based on technical issues and solutions, only 12.5% was about policy and recommendations while 37.5% was about overview and others.

## 4 Conclusion

Non-fungible tokens, also known as NFTs, represent cryptographic properties on the blockchain which include distinct authentication codes as well as metadata which separate them from one another.They can't be sold or swapped in the same way that cryptocurrencies can. This is in contrast to fungible tokens, such as cryptocurrencies, that are all the same and may thus be used as a means of exchange. It represents blockchain-based cryptographic tokens which can never be duplicated. It is possible to utilize NFTs to represent real objects, such as artwork as well as real-estate.As a result of this "tokenization," such real-world tangible goods may be exchanged more quickly, with less risk of corruption. Additionally, NFTs may be used to represent people's personal and financial assets, such as their identities, property ownership, etc. In this work, we conducted a study of eight publications in order to determine which NFT implementations have already been completed and how we may incorporate them into our future technologies.

Furthermore, this survey research has so far included nine subdomains and three large domains, and we hope to work on more domains in the future so that people can use NFT in a variety of domains. This paper's primary objective is to contribute to the latest trending technology and how we can inform others about it for future research. New technologies will always emerge over time, and we wanted to teach how to apply them in a more effective manner for the benefit of humanity.